\documentclass[%
 reprint, 
superscriptaddress,
nofootinbib,
 amsmath,amssymb,
 aps, physrev
]{revtex4-2}

\usepackage{graphicx}
\usepackage{dcolumn}
\usepackage{booktabs}
\usepackage{bm}
\usepackage{physics}
\usepackage{tikz}
\usetikzlibrary{arrows.meta, positioning}

\usepackage{hyperref}

\begin{document}

\preprint{APS/123-QED}

\title{\textbf{Improving the efficiency of Hartree--Fock--Bogoliubov solvers in 3D space} 
}%

\author{A. Sala}
\affiliation{Dipartimento di Fisica “Aldo Pontremoli”, Università degli Studi di Milano, 20133 Milano, Italy}
\author{G. Colò}%
\affiliation{Dipartimento di Fisica “Aldo Pontremoli”, Università degli Studi di Milano, 20133 Milano, Italy}
\affiliation{INFN, Sezione di Milano, 20133 Milano, Italy}

\date{\today}

\begin{abstract}
The solution of the three-dimensional Schrödinger-like single-particle equations that appear in Kohn Sham density functional theory, as well as in other contexts, for large systems and without any symmetry, requires efficient and robust numerical algorithms. Conventional methods suffer from slow convergence and require careful tuning, depending on the spatial discretization.
Conjugate gradient methods combined with preconditioning have been proposed to accelerate the convergence of symmetry-unrestricted Skyrme energy density functionals; however, their effectiveness may depend on the design of a preconditioner. In this work, we introduce the generalized conjugate gradient method for the self-consistent solution of the Hartree--Fock--Bogoliubov equations, which eliminates the need for problem-dependent preconditioning and improves the convergence speed of currently available methods.
The performance of the proposed algorithm is demonstrated on representative nuclear systems, showing improved convergence behavior compared to standard approaches.
The proposed method ultimately provides a promising tool for systematic studies of superheavy, strongly deformed, and drip-line nuclei.
\end{abstract}

\maketitle

\section{Introduction}
Density functional theory (DFT) has been used as a tool to describe nuclear structure since the earliest developments of nuclear theory~\cite{Schunck_DFT,ndft}.
It is one of the many-body techniques capable of handling heavy nuclei without significant computational barriers.
DFT allows to calculate many properties of interest like ground states, potential energy surfaces~(PESs) and static fission paths. It provides the foundation for time-dependent extensions as well, such as time-dependent Hartree--Fock~\cite{sky3d_1p2} and quasiparticle random phase approximation~(QRPA)~\cite{nmbp}.
Beyond mean-field methods, meaning the mixing of different ground state configurations, can be accessed as well~\cite{BMFT_review}. 
Although the spirit of DFT is to describe the quantum systems only by the one-body particle density~\cite{HK}, in most cases nuclear practitioners used the Kohn--Sham method~\cite{KS} and resort to single-particle orbitals to formulate energy functionals that can accurately describe finite nuclei. 
At the core of the problem is the minimization of an energy functional with respect to the single-particle orbitals, which can be formulated as the repeated diagonalization of a single-particle Hamiltonian.
For a complete description of open-shell nuclei, one has to include pairing correlations as well, which result in the Hartree--Fock--Bogoliubov~(HFB), or Bogoliubov--de~Gennes equations~\cite{KS_HFB}.
There are many DFT formulations for finite nuclei, most notably: Gogny, relativistic and Skyrme energy density functionals (EDFs). In this work we only focus on Skyrme EDFs.

There are two classes of methods that are typically used to numerically represent and solve the problem: either using a basis, usually the Harmonic Oscillator basis~\cite{hfodd, hfbtho}, or directly in coordinate representation, by discretizing a finite spatial domain.
The basis approach generally requires considering a truncated basis; although efficient, its complexity scales badly whenever spatial containment of the nucleus, or the cutoff in momentum space, becomes a problem, such as in heavy, very-deformed or drip-line nuclei.
While different bases can be used depending on the pyhsical system, the numerical accuracy is tied to the optimization of the basis parameters; moreover, the computational cost of solving the problem deteriorates rapidly when symmetry-breaking bases are used.
Coordinate representation, on the other hand, has been used since the earliest applications of nuclear EDFs. In particular, it has been used with Skyrme EDFs, where the functional is written in terms of local densities only~\cite{early_skyrme}. Early applications of such method used spherical symmetry to reduce the dimensionality of the problem to one treatable with old computers. In recent years, efforts to push the method to 2D (axial)~\cite{hfbax}, and 3D Cartesian meshes~\cite{ev8,two_basis_hg_bands,cr8,covariant_lobpcg,hfbfft,mixed_basis} have been made. The latter allows the treatment of symmetry breaking configurations, exotic and strong deformations without hindering the numerical accuracy.
The drawback of the full lattice space is that the basis size, i.e.~the number of points on the mesh, is very large, exceeding $10^{5}$ for most applications.
The iterative self-consistent solution in this case cannot be found by exact diagonalization of the single-particle matrix, which is done in all other applications, as the time required to do so is too large. Instead, more advanced iterative algorithms have to be used and have been developed.

A common algorithm used in early applications of 3D meshes is the imaginary time evolution (ITE)~\cite{ite_static}. It is a steepest descent method applied to the minimization of the energy functional. Although simple, this algorithm is slow and has bad scaling with respect to the mesh spacing.
Recent works proposed the use of preconditioning~\cite{ryssens_numerical}, also in combination with the locally optimal conjugate gradient (LOBPCG) method~\cite{zr110}, increasing the complexity of the algorithm, but drastically reducing the iterations needed to reach convergence.
In this work, we propose an improvement to preconditioned conjugate gradients that is based on the generalized conjugate gradient (GCG) method explained in Refs.~\cite{gcg1,gcg2} for the diagonalization of large-scale matrices. This is achieved by changing the preconditioning of the residuals with the use of an inverse Hamiltonian step. This effectively removes the need to choose and design a preconditioner that is problem-specific. The proposed method is also simple and devoid of numerical parameters that need to be tuned for optimal convergence, making it ideal for similar applications as well.
Although the inverse Hamiltonian has been proposed and used for EDFs minimizations, it has never been applied together with conjugate gradient techniques. Moreover, it improves the performance of 3D EDF applications by converging in less iterations and CPU time than other algorithms proposed in the literature to the best of our knowledge.
The introduction of efficient iterative methods allows the systematic application of Cartesian based methods not only to light and medium mass nuclei, but also to superheavy, strongly deformed, drip-line, and symmetry breaking systems as well, where previously the limitation was either numerical error in basis methods, or the high computational demand of the full 3D lattice space.

This article is organized as follows: first, we review the main ingredients and theoretical framework of the Skyrme EDF in Sec.~\ref{sec:theory}. Then, we review the numerical details of a mesh calculation and we introduce the GCG method in Sec.~\ref{sec:numerical}. Finally, we compare the results obtained by our new method for $^{240}$Pu with the ones obtained using the 3D Cartesian code MOCCa~\cite{ryssens_precision} in Sec.~\ref{sec:applications}.
We also showcase the efficiency of the method in exploring octupole deformations in a systematic way by studying the total energy of $^{224}$Ra.

\section{\label{sec:theory}Theoretical framework}
In this section we proceed by introducing the theoretical framework used in this work.
In Sec.~\ref{sec:skyrme} we repeat, just for completeness, the standard formulation for a Skyrme EDF.
The same is done in Sec.~\ref{sec:pairing} where we introduce pairing correlations in the Skyrme EDF by using a zero-range density-dependent pairing interaction.
In Sec.~\ref{sec:two_basis} we explain one of the methods to solve the HFB problem that we have used in this work, namely the two-basis method. Finally, in Sec.~\ref{sec:def} the deformation parameters that are present in this work are briefly sketched to fix the convention used.
\subsection{\label{sec:skyrme}Skyrme EDF}
The total energy of the system can be expressed using an EDF by accounting for five different contributions:
\begin{align}
\label{eq:functional}
E&= E_\text{kin} + E_\text{Sk} + E_\text{cm}+E_\text{Coul}+E_\text{pair}\\
&= \int d\boldsymbol{r}[\mathcal E_\text{kin} + \mathcal E_\text{Sk} + \mathcal E_\text{cm}+\mathcal E_\text{Coul}+\mathcal E_\text{pair}],
\end{align}
where $E_\text{kin}$ is the kinetic energy, $E_\text{Sk}$ the Skyrme energy, which models the strong interaction between the nucleons, $E_\text{cm}$ the center or mass correction, $E_\text{Coul}$ the Coulomb interaction due to the protons and $E_\text{pair}$ the interaction describing pairing correlations.

\subsubsection{\label{sec:e_sk}Skyrme energy functional}
The Skyrme part of the EDF models the strong interaction between the nucleons. From this point onward we will restrict ourselves to the case of conserved time-reversal symmetry. In such case, time-odd densities vanish, and the Skyrme energy reduces to~\cite{unedf, chabanat1, chabanat2}
\begin{eqnarray}
\nonumber
\mathcal E_\text{Sk} = \sum_{t=0,1}\bigg\{&C_t^{\rho}[\rho_0]\rho_t^2 +C_t^{\rho_t}\tau_t + C_t^{\Delta\rho}\rho_t\nabla^2\rho_t+\\\label{eq:skyrme}& +C_t^{\nabla\cdot \boldsymbol{J}}\rho_t\nabla\cdot\boldsymbol{J}_t + C_t^{\boldsymbol{J}^2}\boldsymbol{J}^2\bigg\},
\end{eqnarray}
where the index $t$ runs over $0$ for isoscalar densities and $1$ for isovector densities, respectively $\rho_0 = \rho_n + \rho_p$, $\rho_1 = \rho_n - \rho_p$ and similarly for all other quantities. $p$ and $n$ refer to protons and neutrons, respectively. The $C_t^\rho[\rho_0]$ coupling is further split into two contributions
\begin{equation}
  C_t^{\rho} = C_{t0}^\rho+C_{tD}^\rho \rho_0^\sigma.
\end{equation}
The different densities are defined starting from the density matrix in coordinate space 
\begin{equation}
  \rho_q(\bm r\sigma, \bm r'\sigma') = \expval{a_q^\dagger(\bm r\sigma)a_q(\bm r'\sigma')},
\end{equation}
where the index $\sigma=\pm 1$ runs over the up ($\sigma = 1$) and down ($\sigma = -1$) components of the spin.
The spin trace of the density matrix, which is the total density, reads
\begin{align}
  \rho_q(\bm r, \bm r') = \sum_{\sigma =\pm 1}\rho_q(\bm r\sigma, \bm r'\sigma'),
\end{align}
and the spin density reads
\begin{equation}
  \bm s_q (\bm r, \bm r')=\sum_{\sigma\sigma' = \pm 1}\rho_q(\bm r, \bm r')\bra{\sigma'}\bm \sigma \ket{\sigma'},
\end{equation}
where $\bm \sigma$ is the vector of Pauli matrices.
The densities in Eq.~\eqref{eq:skyrme} are then defined as
\begin{align}
\rho_q &= \rho_q(\bm r, \bm r')|_{\bm r = \bm r'},
\\ \tau_q &= \nabla'\cdot\nabla\rho(\bm r, \bm r')|_{\bm r = \bm r'},
\\J_{q, \mu\nu} &= \frac{1}{2i}(\nabla_\mu - \nabla_\mu') s_{q, \nu}(\bm r, \bm r')|_{\bm r = \bm r'},
\\J_{q,\kappa}(\boldsymbol{r})&=\sum_{\mu\nu}\varepsilon_{\kappa\mu\nu}J_{q,\mu\nu}\label{eq:J_vec},
\end{align}
where $\varepsilon_{\kappa\mu\nu}$ is the Levi-Civita symbol.
The spin-orbit density current tensor $J_{\mu\nu}$ contribution to the EDF is generally split into its irreducible components.
The scalar and tensor (irreps) contributions vanish in our time-reversal invariant description. The only non-zero contribution is given by the time-odd vector component $J_{q,\kappa}$ defined in Eq.~\eqref{eq:J_vec}.
The explicit expression of the densities in coordinate representation is given in App.~\ref{app:densities}.
\subsubsection{Kinetic energy and center of mass correction}
The kinetic energy term reads
\begin{equation}
  \mathcal E_\text{kin} = \frac{\hbar^2\tau}{2m}.
\end{equation}
As translational symmetry is broken when using a Slater determinant, this breaking is approximately accounted for by subtracting the one-body center of mass correction correction term, which reads
\begin{equation}
  \mathcal E_\text{cm} = -\frac{1}{2mA}\sum_i \expval{p_i^2}.
\end{equation}
This equates to a renormalization of the nucleon mass, which then takes the form
\begin{equation}
  m\to m\frac{A}{A-1}.
\end{equation}
\subsubsection{Coulomb energy}
The Coulomb energy is treated within the Local Density Approximation. The resulting energy density reads~\cite{SlaterApp}
\begin{align}
    \mathcal E_\text{Coul} &=\mathcal E_{\text{Coul, D}}+\mathcal{E}_\text{Coul, E} \\&=\frac{e^2}{2}\bigg[\int U_\text{c,D}(\boldsymbol r)\rho_p(\boldsymbol r)d\boldsymbol r    - \frac 3 2 \bigg(\frac 3 \pi \bigg)^{\frac 1 3}\rho_p^{4/3}(\bm r)\bigg],\nonumber
\end{align}
where the Coulomb field is the one generated by the protons. The details on how the field is computed are given in App.~\ref{app:coulomb}.

\subsubsection{Pairing energy}
The pairing interaction is assumed to be different from the particle-hole one, as done extensively in the literature.
The pairing interaction when using a Skyrme functional is often taken to be zero-range. Discussion on the differences provided by a zero- or a finite-range force can be found in the literature. Here, we carry on with a standard formulation.
Treating the particle-particle channel requires the definition of the skew-symmetric pairing tensor
\begin{equation}
\kappa_q (\bm r \sigma, \bm r'\sigma') = \expval{a_q(\bm r'\sigma') a_q(\bm r \sigma)},
\end{equation}
which can be cast on a pairing density matrix that reads
\begin{equation}
\tilde\rho_q(\bm r \sigma, \bm r'\sigma')=-\sigma'\kappa_q(\bm r \sigma, \bm r'-\sigma').
\end{equation}
The local pairing density finally reads 
\begin{equation}
  \label{eq:pairing_density}
\tilde\rho_q(\bm r)=\sum_{\sigma=\pm 1}\tilde\rho_q(\bm r \sigma, \bm r\sigma).
\end{equation}
In this work, we employ a zero-range interaction which results in the energy density~\cite{Bender2003}
\begin{equation}
    \mathcal E_\text{pair} =\sum_{q=n,p} \frac 1 4 V_{q}\tilde\rho_q(\bm r)\tilde\rho_q^*(\bm r)\bigg[1-\eta\frac{\rho_0(\bm r)}{\rho_s}\bigg].
\end{equation}
Here, $\eta$ is a parameter which controls the spatial shape of the pairing interaction, $\eta=0$ corresponds to volume pairing, $\eta=1$ to surface pairing, and mixed pairing is usually taken with $\eta=1/2$. $\rho_s$ is another free parameter, usually taken as the nuclear saturation density, fixed at $0.16$ fm$^{-3}$.
\subsection{\label{sec:pairing}HFB equations}
Minimizing the EDF in the Kohn--Sham framework~\cite{KS}, by constraining the single-particle wavefunctions orthonormality and the particle number average, leads to the Hartree--Fock--Bogoliubov, or Bogoliubov--de Gennes~\cite{KS_HFB} equations
\begin{equation}
  \begin{pmatrix}
    h - \lambda & \Delta \\ -\Delta^* & -(h^*-\lambda)
  \end{pmatrix}
  \begin{pmatrix}
    U_\mu \\ V_\mu
  \end{pmatrix}
  =E_\mu \begin{pmatrix}
    U_\mu \\ V_\mu
  \end{pmatrix},
  \label{eq:HFB}
\end{equation}
where $E_\mu$ are the quasiparticle energies and $U_\mu$ and $V_\mu$ are, respectively, the upper and lower components of the quasiparticle wavefunctions. The mean-field Hamiltonian $h$ and the pairing field $\Delta$ are obtained as variations of the functional, respectively $h_{ij} = \delta \mathcal E / \delta \rho_{ij}$ and $\Delta_{ij} = \delta \mathcal E / \delta \kappa^*_{ij}$. 
$\lambda$ is a Langrange multiplier used to fix the particle number on average, as the quasiparticle vacuum does not conserve the particle number by breaking the $U(1)$ global symmetry. 

\subsubsection{Mean-field Hamiltonian}
The variation of the energy functional with respect to the density matrix $\rho_{ij}$ leads to the single-particle Hamiltonian, which projected on the coordinate basis reads
\begin{equation}
h=\nabla\cdot\frac{\hbar^2}{2m^*(\boldsymbol{r})}\nabla+U_{q}(\boldsymbol r)+\boldsymbol B_q(\boldsymbol r)\cdot(\nabla\times\hat\sigma),
\end{equation}
where the different fields in the equation are computed from the variation of the energy functional with respect to the local densities
\begin{equation}
  \label{eq:fields}
U_q = \frac{\delta \mathcal E}{\delta \rho},\ \boldsymbol B_q = \frac{\delta \mathcal E}{\delta \boldsymbol{J}},\ \frac{\hbar^2}{2m^*}=\frac{\delta \mathcal E}{\delta \tau}.
\end{equation}
The self-consistent solution of Eq.~\eqref{eq:HFB} in coordinate space can be obtained in different ways. In this work, we employ the two-basis method; a detailed discussion can be found in Refs.~\cite{cr8,TERASAKI19951,two_basis_hg_bands}.
\subsection{\label{sec:two_basis}Two-basis method}
In the two-basis method, instead of diagonalizing the full HFB matrix, the problem is split into finding two bases: (a) the HF basis which diagonalizes the mean-field Hamiltonian $h$ and (b) the quasiparticle basis, expressed in the HF basis, which diagonalizes the HFB Hamiltonian.
In practice, this means that at each iteration we solve the eigenvalue problem
\begin{equation}
  h\ket{\phi_k} = \varepsilon_k\ket{\phi_k},
\end{equation}
which yields the HF basis $\ket{\phi_k}$. We can then compute the local pairing density in Eq.~\eqref{eq:pairing_density} as 
\begin{equation}
  \tilde\rho_q(\bm r) = \sum_{\sigma=\pm 1}\kappa_{ij}f_i f_j\phi_i(\bm r\sigma)\phi_j(\bm r -\sigma),
\end{equation}
and the pairing gaps in this basis as
\begin{equation}
\Delta_{ij}=\sum_{\sigma=\pm 1}\int d\bm r \frac{V_q} 2 \bigg[1-\eta\frac{\rho_0(\bm r )}{\rho_s}\bigg]\phi_i(\bm r \sigma)\phi_j(\bm r -\sigma)f_i f_j.
\end{equation}
The $f_i,f_j$ factors are used to regularize the pairing interaction, as $\mathcal E_\text{pair}$ is local in the pairing density, implying constant coupling in momentum space.
Without some kind of regularization, the pairing energy for a zero-range interaction would depend on the number of basis states and diverge in the limit of infinite basis states. There are different strategies one can use, and in this work we exclude all states outside the pairing window.
To avoid sudden changes in the states considered, the factors used are the smooth ones proposed in Ref.~\cite{pairing_factors}, which read
\begin{equation}
  \nonumber
  f_i = \bigg[\frac 1 {1+e^{(\varepsilon_i-\lambda_q-\Delta\varepsilon_q)/\mu_q}}\cdot\frac 1 {1+e^{(-\varepsilon_i+\lambda_q-\Delta\varepsilon_q)/\mu_q}}\bigg]^{1/2},
\end{equation}
where $\Delta\varepsilon_q$ is the pairing window and $\mu_q$ is a factor which controls the sharpness of the window.
After diagonalizing the HFB Hamiltonian in this basis, the density matrix and the pairing tensor are computed as
\begin{equation}
\rho = V^*V^T,\quad \kappa = U^*V^T.
\end{equation}
The next single-particle Hamiltonian is computed from the densities calculated in the canonical basis $\ket{\psi_i}$, i.e.~the set of orbitals that diagonalizes the density matrix $\rho$
\begin{align}
 (S^{-1}\rho S)_{ij} =  \delta_{ij} v_i^2,
\end{align}
$S$ is the unitary matrix that transforms the HF basis to the canonical basis, which reads
\begin{equation}
  \ket{\psi_i} = \sum_{k}S_{ki} \ket{\phi_k}.
\end{equation}
\subsection{\label{sec:def}Deformation parameters}
The nuclear density can be described in terms of dimensionless deformation parameters.
Beyond spherical symmetry, the lowest order of nuclear deformation is a quadrupole one. In this case, the relevant deformation parameters are
\begin{equation}
  \beta_2 = \frac{4\pi}{3AR^2}\expval{ {Q}_{20}},
\end{equation}
where $ Q_{\lambda\mu}$ is the quadrupole moment operator associated to the spherical harmonics $Y_{\lambda\mu}$ of degree $\lambda$ and order $\mu$,
\begin{equation}
   Q_{\lambda\mu} = r^\lambda Y_{\lambda\mu},
\end{equation}
and
\begin{equation}
  \gamma =\atan\{{\sqrt 2 {\expval{{Q}_{22}}}/{\expval{ Q_{20}}}} \}.
\end{equation}
$R$ is the mean nuclear radius that can be derived from the mean square radius
\begin{equation}
  R = \frac 5 3 \sqrt{\expval{r^2}}
\end{equation}
or by using a suitable parametrization. In this work we use the latter prescription, where the radius reads
\begin{equation}
  R = 1.2A^{1/3}.
\end{equation}
Quadrupole deformations are uniquely defined by restricting $\gamma$ to the interval $[0\text{°},60\text{°}]$ and assigning positive and negative values of $\beta_2$ to prolate and oblate shapes, respectively.
For higher orders of deformation we limit ourselves to axially symmetric shapes, which can be described by
\begin{equation}
  \beta_l = \frac{4\pi}{3AR^l}\expval{Q_{l0}}.
\end{equation}

\section{\label{sec:numerical}Numerical treatment}
In this section, the details of the numerical solution of the HFB equations are given. We start by reviewing the parameters of a Cartesian mesh in Sec.~\ref{sec:mesh} and the prescription to compute the derivatives. In Sec.~\ref{sec:gcg} we explain how to obtain the HF basis at each iteration by implementing the GCG algorithm. Some discussion on the convergence rate of the algorithm is also presented. Finally, in Sec.~\ref{sec:alm} we discuss the implementation of spatial constraints in the self-consistent procedure, along with a novel strategy to improve the convergence speed.
\subsection{\label{sec:mesh}Cartesian mesh and derivatives}
In this work, we use cubic meshes defined by the spatial extent in each direction $[-a, +a]$ and a number of points $N$. The mesh step size $\delta$ is given by $\delta=2a/(N-1)$. When some reflection across a given plane can be assumed, e.g. in spherical, axial or triaxial nuclei, only half length of some sides can be considered. Here, for the sake of simplicity and generality, we use the full lattice space. Thanks to the efficiency provided by the GCG method, results can be obtained in reasonable CPU times even without such symmetry assumptions.

Derivatives that enter the densities at each iteration are computed assuming a Lagrange mesh interpolation~\cite{ryssens_precision}. By using the information of the whole axis it enables reaching accuracies of the keV order for step sizes of $0.5$ fm. Moreover, it guarantees that calculations are variational.
The single-particle Hamiltonian $h$ differential operators are represented in matrix form with a five point finite difference derivative in each direction~\cite{finite_diff}. This is done to reduce the size of the matrix and increase the speed of the inverse application. 

\subsection{\label{sec:gcg}Implementation of the GCG method}
Finding the HF basis amounts to solving the minimization problem
\begin{equation}
  \min_{\{\phi\}} E[\phi] = \min_{\{\phi\}}\frac{\bra\phi h \ket\phi}{\bra{\phi} \ket{\phi}},
    \label{eq:energy_orb}
\end{equation}
which results in the eigenvalue problem
\begin{equation}
  h\ket{\phi_k} = \varepsilon_k\ket{\phi_k}.
  \label{eq:eigenvalue}
\end{equation}
The Lagrange multipliers $\varepsilon_k$, which can be interpreted as the single-particle energies, are used to ensure the orthonormality of the orbitals.
Calculating the HF basis on a cartesian mesh by repeatedly solving Eq.~\eqref{eq:eigenvalue} is unfeasible. Even very modest meshes lead to large-scale matrices which cannot be diagonalized exactly in reasonable CPU times. There exists a wide range of methods that can be employed for this kind of problem.

A common minimization method used is the imaginary time evolution (ITE)~\cite{two_basis_hg_bands,ite_tf,ev8}. The algorithm is quite simple, it starts from an orthonormal set of guess orbitals $|\phi^{(0)}\rangle$ which are evolved at each iteration as
\begin{equation}
  |\phi_k^{(i+1)}\rangle = \bigg(1-\frac{\delta t}{\hbar}h\bigg)|\phi_k^{(i)}\rangle,
\end{equation}
after which they have to be orthonormalized. 
Although derived from the first order expansion of the imaginary time evolution of the Hamiltonian, ITE can be reduced to a simple gradient descent. Starting from the orbitals at iteration $i$, following the steepest descent of the energy functional would give:
\begin{equation}
  |\phi_k^{(i+1)}\rangle  = (1-\alpha h)|\phi_k^{(i)}\rangle,
\end{equation}
with $\alpha>0$ being a parameter tuned to reach the minimum energy.
The attractiveness of the method lies in its simplicity and computational efficiency, only requiring a sparse matrix multiplication and an orthonormalization. The downside is that $\alpha$ ($\delta t$) is a parameter that needs to be tuned for optimal convergence and stability. 
Writing $|\phi_k^{(i)}\rangle$ in terms of the eigenvectors $\varphi_k$ of $h$, the gradient descent reads
\begin{equation}
\sum_k v_k^{(i+1)} |\varphi_k\rangle = (1-\alpha h)\sum_k v_k^{(i)} |\varphi_k\rangle,
\end{equation}
projecting on the $\ell$-th eigenvector, we get an expression for the coefficient
\begin{equation}
  \label{eq:ite_coeff}
  v_\ell^{(i+1)} = (1-\alpha \varepsilon_\ell) v_\ell^{(i)} = (1-\alpha\varepsilon_\ell)^i v_\ell^{(0)}.
\end{equation}
Assuming, for simplicity, a positive energy spectrum, the coefficients $v_\ell^{(i+1)}$ do not diverge if 
\begin{equation}
  0 < \alpha < \frac 2 \varepsilon_\text{max}.
\end{equation}
This means that $\alpha$ has an upper bound determined by the largest eigenvalue of $h$. If we assume to set $\alpha = 1/\varepsilon_\text{max}$, the coefficients at iteration $i+1$ will be given by
\begin{equation}
  v_\ell^{(i+1)} = \bigg(1-\frac{\varepsilon_\ell}{\varepsilon_\text{max}}\bigg)^i v_\ell^{(0)}.
\end{equation}
Here, the slowest coefficient to converge is for $\ell =0$, meaning that the speed of conergence is determined by the ratio $\varepsilon_\text{max}/\varepsilon_0$, a quantity known as condition number $\kappa$ of the matrix $h$~\cite{ryssens_numerical}, that is
\begin{equation}
  \kappa(h) = \bigg|\frac{\varepsilon_{\text{max}}}{\varepsilon_0}\bigg|,\ \kappa(h) \in [1,\infty).
\end{equation}
Here, $\varepsilon_0$ is the smallest eigenvalue (in magnitude) of $h$ and $\varepsilon_{\text{max}}$ is the largest. While the lowest eigenvalue of $h$ is independent of the mesh step size, the largest is inversely proportional to the mesh spacing $\varepsilon_\text{max} \sim 1/\delta$~\cite{ryssens_numerical}.
Although there are empirical formulae that automatically find the optimal value of $\alpha$ for a given mesh spacing~\cite{ryssens_numerical}, the method still suffers from slow convergence and scalability with the decrease of $\delta$.
A promising technique to solve this issues is the use of preconditioning~\cite{ryssens_numerical, covariant_lobpcg}. It is a technique which aims at damping high energy modes to decrease the number of iterations, by transforming the problem into one where $\kappa$ approaches $1$. It can either be done on the potentials~\cite{ryssens_numerical}, on the residuals~\cite{LOBPCG}, or on orbitals themselves~\cite{Saad2003}.
Preconditioning solves the problem of a high condition number, however, the convergence rate is still linear in $1/\kappa$. Defining the error $e_k^{(i)}=|\phi_k^{(i+1)}-\phi_k^{(i)}|$, by some trivial algebra we can write
\begin{equation}
  (e_k^{(i)})^2 = \sum_\ell |v_\ell^{(i+1)}-v_\ell^{(i)}|^2 = \sum_\ell \bigg(1-\frac{\varepsilon_\ell}{\varepsilon_\text{max}} \bigg)^2|v_\ell^{(i)}|^2.
\end{equation}
For large values of $i$, all components are negligible with respect to the slowest one, i.e.~$\ell=0$. We can thus approximate the error as
\begin{equation}
  e_k^2 \lesssim \bigg(\frac{\kappa-1}{\kappa}\bigg)^2.
\end{equation}
We can then write, for large $\kappa$:
\begin{equation}
  e_k \lesssim \bigg|\frac{\kappa-1}{\kappa}\bigg| \sim \bigg|1 - \frac 1 \kappa\bigg|.
\end{equation}

This bound can be improved by using a \textit{conjugate gradient}; an in-depth explaination can be found in Ref.~\cite{painless_cgm}. Without going into the details, the idea is to retain memory of the past searches by taking steps in subspaces of the vector space orthogonal to the previously explored ones. Expressing the new search step as 
\begin{equation}
|d^{(i)}\rangle=|\phi^{(i)}\rangle -|\phi^{(i-1)}\rangle,
\end{equation}
it can be proven for a quadratic objective function that if every step is $h$-conjugate to the previous one, meaning
\begin{equation}
  \langle d^{(i-1)} | h | d^{(i)} \rangle = 0
\end{equation}
then after $n$ iterations the full vector space is explored, where $n$ is the number of rows (columns) of the square matrix $h$.
This is rather inexpensive from a computational point of view, but it can be shown that the error dependence on $\kappa$ changes as
\begin{equation}
  e_k \lesssim \bigg|1- \frac{1}{\sqrt \kappa}\bigg|.
\end{equation}
Algorithms based on these three ideas, namely gradient descent, conjugate search directions, and preconditioning, are calleded preconditioned conjugate gradient methods and are generally formulated for the minimization of quadratic forms 
\begin{equation}
  f(\phi) = \frac 1 2 \langle\phi| h |\phi\rangle - \langle b| \phi\rangle,
\end{equation}
where $b$ is a vector of known values, $h$ a fixed matrix, and the pedix for $\phi$ is omitted to indicate a matrix of orbitals. In such cases, the convergences of the method is exact for a finite number of iterations. However, in the case of the HFB equations, and more specifically in the two-basis method, the function to be minimized takes the form of a Rayleigh-Ritz quotient:
\begin{equation}
  \label{eq:obj}
  f(\phi) = \frac{\bra \phi h \ket \phi}{\bra \phi \ket{\phi}} ,
\end{equation}
which makes the optimization non-linear. Moreover, in HFB, $h$ depends on the densities. Some algorithms have been developed in recent years to tackle this problem in the preconditioned conjugate gradient framework when the matrix is constant and its size is large, such as the locally optimal preconditioned conjugate gradient (LOBPCG)~\cite{LOBPCG} and the generalized conjugate gradient (GCG)~\cite{gcg1, gcg2}.
In this work we implement an algorithm based on the GCG method, where preconditioning is replaced by an inverse Hamiltonian step~\cite{hfb_ihm}.
A short review of preconditioning and the inverse Hamiltonian is given in App.~\ref{app:preconditioning}.

\subsubsection{Algorithm}
The method starts with an ansatz of $N_b$ single-particle wavefunctions $\{\phi_k^{(0)}\}$ and iteratively updates them at each mean-field iteration, until some convergence criterion is met.
At iteration $(i)$ of the calculation, a matrix of column vectors $W$ is constructed by an application of the inverse single-particle hamiltonian $h^{(i)}$, solving the linear system
\begin{equation}
  \label{eq:inv_sys}
    (h^{(i)}-\varepsilon_{0,s}^{(i)})W^{(i)}_k = (\varepsilon^{(i)}_k-\varepsilon_{0,s}^{(i)})\phi_k^{(i)},
\end{equation}
where $\varepsilon_{0,s}^{(i)}$ is the lowest single-particle energy $\varepsilon_0$, at iteration $(i)$, shifted by a small quantity $c>0$, which can be taken as 
\begin{equation}
c=|\varepsilon_0^{(i)}|/100.
\end{equation}
The solution of the linear system in Eq.~\eqref{eq:inv_sys} can be obtained by a wide variety of iterative algorithms, available in most modern linear algebra libraries.
A conjugate search direction column vector matrix is built using the previous iteration's states
\begin{equation}
P_k^{(i)} = \phi^{(i)}_k - \langle\phi_k^{(i-1)}| \phi_k^{(i)}\rangle\phi_k^{(i-1)},
\end{equation}
where $P_k^{(0)}=0$. 
The new states for the iteration $(i+1)$ are computed using the Rayleigh-Ritz method~\cite{RR_method}, which corresponds to taking the optimal step $\alpha$ in the vector space 
\begin{equation}\mathcal V^{(i)}=\text{span}\{W^{(i)}, P^{(i)}, F^{(i)}\},
\end{equation}
where $F^{(i)}$ is the column vector matrix 
\begin{equation}
F_k^{(i)} = \phi_k^{(i)}.
\end{equation}
This is achieved by solving the eigenvalue problem
\begin{equation}
  V^{(i)\dagger} h^{(i)} V^{(i)}C = \Lambda C
\end{equation}
which is of dimension $3N_b\times 3N_b$. $V^{(i)}$ is a matrix constructed by building the column vector matrix $[W^{(i)}, P^{(i)}, F^{(i)}]$ and orthonormalizing it. $\Lambda$ is a diagonal matrix containing the eigenvalues of the small eigenvalue problem and $C$ is a matrix containing the eigenvectors.
The new eigenvectors $\{\phi_k^{(i+1)}\}$ are the ones corresponding to the lowest $N_b$ eigenvalues and they are obtained by
\begin{equation}
  \phi_k^{(i+1)} = \sum_{j=1}^{N_b} \phi_j^{(i)}C_{kj}.
\end{equation}
After obtaining the new HF basis $\{\phi_k^{(i+1)}\}$, the HFB matrix can be diagonalized, after which the canonical basis $\{\psi^{(i+1)}\}$ is computed. With that, the new local densities can be calculated, as well as the single-particle matrix $h^{(i+1)}$. This self-consistent procedure is repeated until convergence.

\subsubsection{Comparison with other algorithms}
\begin{figure}[ht]
  \includegraphics[ width=0.9\columnwidth]{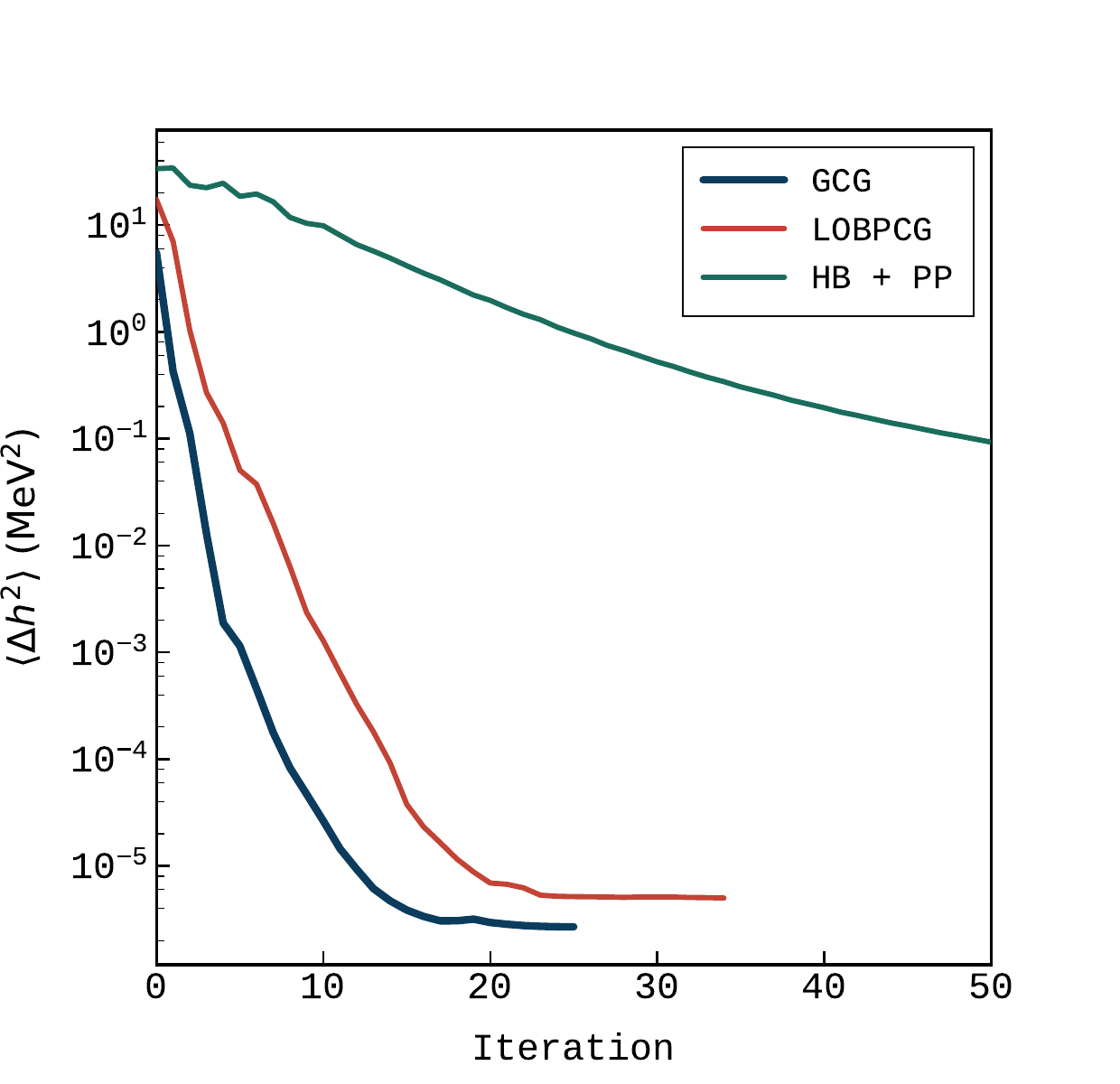}
  \caption{Convergence of a spherical calculation in $^{44}$Ca using SLy5~\cite{chabanat2}, including pairing correlations. The mesh used has a spatial extent of $30\times 30\times 30$~fm and a step size of $1.0$~fm. The time required to reach a weighted dispersion value of $10^{-5}$ MeV$^2$ for GCG, LOBPCG and HB+PP is, respectively, 181~s, 319~s and 630~s on a AMD Ryzen 5 7640U.}
  \label{fig:conv_comp}
\end{figure}

The convergence of a calculation can be assesed in different ways. One can use the average weighted single particle dispersion
\begin{equation}
  \langle{\Delta h^2}\rangle = \sum_{k=1}^{N_b}\rho_{kk} [\langle{\phi_k^{(i)}}|h^2|\phi_k^{(i)}\rangle - (\varepsilon_k^{(i)})^2],
\end{equation}
because this quantity should ideally reach 0, although it is numerically bound by the truncation of the Hilbert space when using a mesh (or a finite basis). It can be used as a criterion to compare different algorithms, as its value is independent on the procedure used.
As shown in Fig.~\ref{fig:conv_comp}, GCG achieves the same dispersion as the very similar counterpart LOBPCG~\cite{LOBPCG} using~40\% less iterations when using the preconditioner
\begin{equation}
  T = \frac{\boldsymbol{\hat p}^2}{2m}, 
\end{equation}
where $\boldsymbol{\hat p}$ is the momentum operator in coordinate space.
Moreover, no design of an optimal preconditioner is required, as the single-particle matrix is used in the inverse step.
The application of a preconditioner requires calculating a product of an inverse matrix, rendering the computational cost, per iteration, of GCG and LOBPCG similar.

Another interesting comparison can be done with the algorithm heavy ball + potential preconditioning (HB+PP) proposed in \cite{ryssens_numerical}. While its computational overhead is very light, being very similar to ITE, and requiring the preconditioning of a single (potential) vector, the execution time is dominated by the recalculation of the derivatives, making the method less efficient than the two conjugate gradient algorithms, since the number of iterations required to reach convergence is much higher, even when the mesh step size is kept at $1.0$ fm.

\subsubsection{Mesh step size and computational scaling}
\begin{figure}
  \includegraphics[ width=0.9\columnwidth]{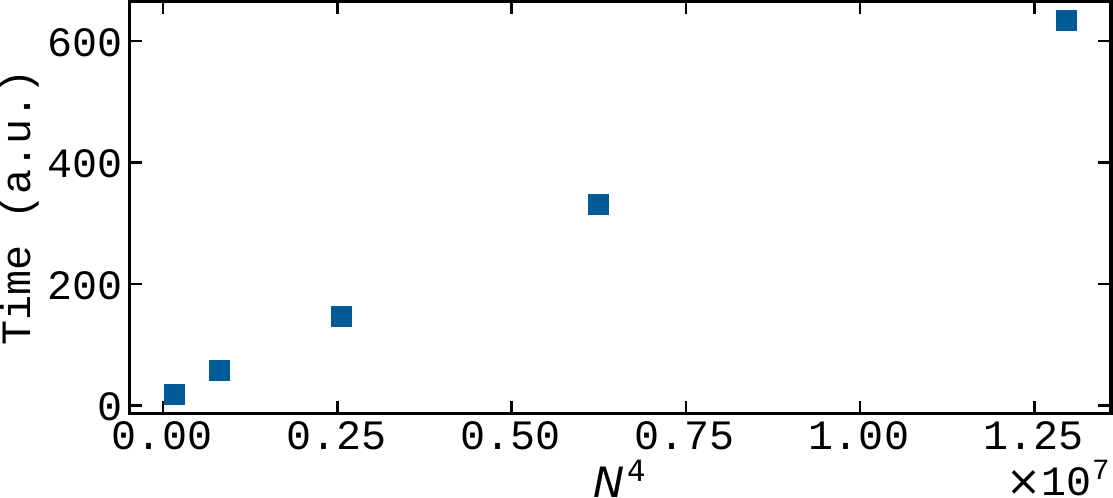}
  \caption{Convergence of a deformed calculation in the $^{20}$Ne system using SLy5~\cite{chabanat2}. The box size is kept fixed at 20 fm for each spatial direction. The step size starts at 1.0 fm and reaches 0.3 fm for the last data point. The full self-consistent calculation appears to scale with a complexity between $O(N^3)$ and $O(N^4)$, where $N$ is the number of mesh points in each direction.}
  \label{fig:N_scaling}
\end{figure}

An interesting feature of the GCG method, and of preconditioning in general, is that the number of iterations needed to reach convergence is independent of the mesh size, as the condition number of the problem approaches $1$. This implies that the computational cost of the lattice only scales with number of points, as highlighted in Fig.~\ref{fig:N_scaling}, and not the step size $\delta$.
The  cost for a generic self-consistent calculation is dominated by the computation of the derivatives and the inverse Hamiltonian step. The scaling for computing a given derivative on a Lagrange mesh along a given axis is $O(N^4)$, as they have to be computed for each point and they need the information of the whole axis. The inverse Hamiltonian step can be solved with an iterative method~\cite{Saad2003}, which in general scales as the number of non-zero elements in the matrix. Since a five-point scheme is used, the scaling is between $O(N^3)$ and $O(N^4)$.

\subsection{\label{sec:alm}Spatial constraints}
\begin{figure}[ht]
  \centering
  \includegraphics[width=0.9\columnwidth]{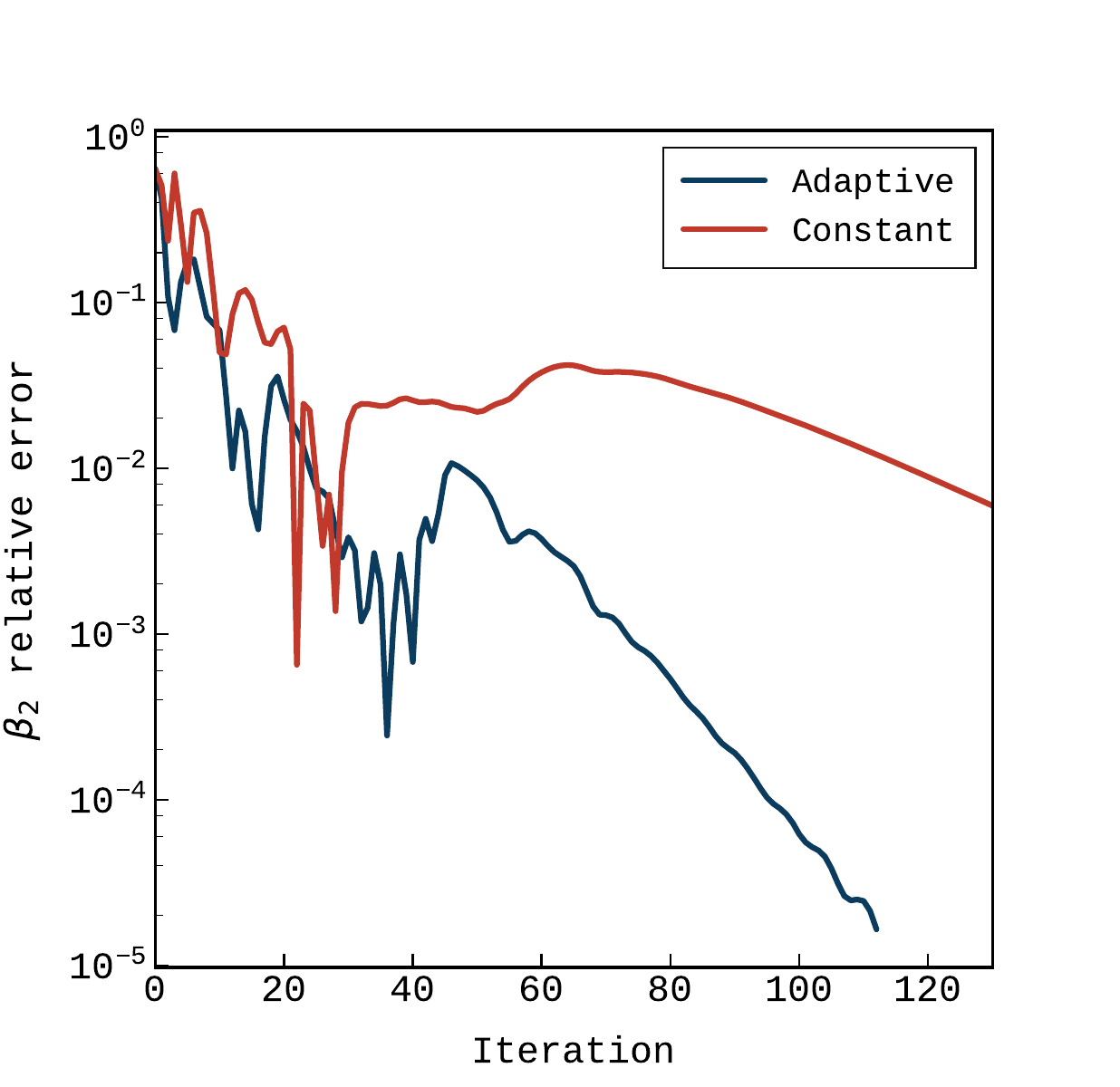}
  \caption{Convergence of the constraint on $\beta_2$ for a constrained calculation of $^{24}$Mg to $\beta_2=0.2$ using the SLy5 functional~\cite{chabanat2}. With \textit{constant} we refer to an update of $\lambda^{(i)}$ at each iteration using $\mu=0.02$, while with \textit{adaptive} we refer to the proposed strategy, updating $\lambda^{(i)}$ only when $\delta Q^{(i)} < \epsilon$, using $\mu = 0.2$.}
  \label{fig:const_conv}
\end{figure}

Constraining multipole moments of the nuclear density is necessary to explore potential energy surfaces (PESs). In practice, this means requiring that the expectation value of an operator $Q$ of choice satisfies
\begin{equation}
  \expval{Q} = q_\text{target} = q
\end{equation}
at convergence.
Standard approaches either use a Lagrange multiplier or a quadratic penalty~\cite{Flocard1973}. The former can achieve exact convergence on the target value; however, the converged state is limited to convex regions of the PES~\cite{Flocard1973}. The latter can explore concave parts of the PES, and is used in modern codes~\cite{hfodd,hfbtho}, but convergence of $\expval Q$ is only approximate and dependent on the constraint stiffness. To solve these issues, spatial constraints are implemented in this work by using the Augmented Lagrangian Method~\cite{ALM}, which minimizes the Routhian
\begin{equation}
E' = E + \lambda^{(i)} (\langle Q \rangle-q) + \frac c 2 (\langle Q\rangle -q)^2,
\end{equation}
where $Q$ is a multipole operator, $q$ a target value for said operator, and  $\lambda^{(i)}$ is a Lagrange multiplier. $\lambda^{(i+1)}$ is computed by
\begin{equation}
    \lambda^{(i+1)} = \lambda^{(i)}+\mu c (\langle Q\rangle -q),
\end{equation}
where $\mu\approx 0.2$ is a constant used to damp oscillations of the constraint field. Instead of the usual strategy of performing a $\lambda$ update at each mean-field iteration~\cite{ev8,mixed_basis}, we propose a more efficient prescription. We update $\lambda$ only when 
\begin{equation}
  \delta Q^{(i)} = \frac{|\expval Q ^{(i-1)} - \expval Q^{(i)}|}{|\expval Q ^{(i-1)}|} < \epsilon,
\end{equation}
where the tolerance $\epsilon$ is set at $0.01$. This prescription is well suited for fast convergence algorithms: an unsupervised constant update is either unstable when $\mu$ is large or slow when $\mu$ is small. The quick convergence of GCG for a fixed $\lambda$ guarantees that it gets updated a sufficient amount of times to convergence. We compare the two methods in Fig.~\ref{fig:const_conv}, where we denote as \textit{adaptive} our proposed strategy and as \textit{constant} the usual update at each mean-field iteration with $\mu=0.02$.

\subsection{Particle number constraint}
The Lagrange multiplier $\lambda$ in Eq.~\eqref{eq:HFB} is used to fix the particle number on average $\expval{N}$. It is determined during the diagonalization of the HFB matrix using a bisection algorithm.
This can be done by first determining a bracketing interval $[\lambda_\text{min}, \lambda_\text{max}]$ in which the correct value of $\lambda$ should lie. The bisection algorithm then proceeds as follows:
\begin{enumerate}
\item Diagonalize the HFB matrix 
\begin{equation}
\begin{pmatrix}
  h - \lambda^{(j)} & \Delta \\ -\Delta^* & -(h^*-\lambda^{(j)})
\end{pmatrix}
\begin{pmatrix}
  U_\mu \\ V_\mu
\end{pmatrix}^{(j)}
=E_\mu \begin{pmatrix}
  U_\mu \\ V_\mu
\end{pmatrix}^{(j)};
\end{equation}
\item determine the particle number $\expval{N}^{(j)}$;
\item if $\expval{N}^{(j)} < A$, set $\lambda_\text{min} = \lambda^{(j)}$; otherwise set $\lambda_\text{max} = \lambda^{(j)}$;
\item set $\lambda^{(j+1)} = (\lambda_\text{min} + \lambda_\text{max})/2$. Repeat until convergence.
\end{enumerate}

The superscript $(j)$ is used to avoid confusion between the outer mean-field iterations $(i)$ and the inner bisection iterations $(j)$. During the bisection algorithm, only the $\lambda^{(j)}$ value is updated, for which a new quasiparticle basis is obtained. All other quantities are kept fixed as to prevent ill-defined convergence of the algorithm.


\section{\label{sec:applications}Applications}
In this section, we demonstrate the performance of the proposed method to solve the HFB equations on two heavy nuclei, $^{240}$Pu and $^{224}$Ra. We first consider $^{240}$Pu in Sec.~\ref{sec:pu}; being a system extensively studied with Skyrme EDFs, we use it to assess the validity of our algorithm. We then apply the method to $^{224}$Ra in Sec.~\ref{sec:ra}, where exploring symmetry-breaking configurations is crucial to correctly describe the ocutpole deformed ground state.

\subsubsection{Numerical details}
To compare our results for $^{240}$Pu with MOCCa~\cite{ryssens_numerical}, for all the calculations from here onwards, we used a surface pairing interaction ($\eta=1$) and a pairing strength $V = V_n=V_p =1250$ MeV$\cdot$fm$^{3}$. The pairing window is set to $\Delta\varepsilon = 5$ MeV and the diffuseness $\mu=0.5$ MeV for both proton and neutron. The Skyrme functional employed is SLy4~\cite{chabanat2}.
All parity breaking calculations have been performed constraining the center of mass to the origin.

\subsection{\label{sec:pu}$^{240}$Pu Fission barrier}
One of the main driver for using meshes over basis expansion methods is that the numerical error does not depend on the nuclear shape~\cite{DOBA_dripline}. This is a major advantage when very deformed shapes need to be studied, as in the case of nuclear fission.
A typical system studied for fission is $^{240}$Pu. Detailed calculations regarding the dynamics of the process have already been carried out in the literature, some examples can be found in Refs.~\cite{pu240_fission,Pu240_fission_fragments,Pu240_fission_langevin,Pu240_induced}. Here, we limit ourselves to a PES in the $\beta_2$ deformation parameter to values up to $~1.5$. We also compare our results with those obtained by the MOCCa code, whose details are given in Ref.~\cite{ryssens_precision}.
In Tab.~\ref{tab:pu240_comparison_gs}, \ref{tab:pu240_comparison_gs_fi} we report the calculated values of total energy and $\beta_2$ of the ground state and superdeformed fission isomer for different meshes. The two codes show good agreement, having total energies which differ by~$\approx 200$ keV and~$\approx 300$ keV for similar step sizes, for the ground state and the fission isomer respectively. The value for $\beta_2$ is in good agreement as well, with differences in the order of $0.5\%$.
\begin{table}[t]
\caption{
Comparison of the calculated ground state properties of
$^{240}$Pu with the results obtained using
MOCCa~\cite{ryssens_precision}.}
\label{tab:pu240_comparison_gs}
\begin{ruledtabular}
\begin{tabular}{lcccc}
 & 
 $L_x \times L_y \times L_z$ (fm) &
 Step (fm) &
 $\beta_{2}$ &
 $E_{\mathrm{tot}}$ (MeV)
\\
\midrule
This work &
$32 \times 32 \times 32$ &
0.681 &
0.289 &
$-1802.763$ 
\\
This work &
$40 \times 40 \times 40$ &
0.678 &
0.289 &
$-1802.768$ 
\\
This work &
$32 \times 32 \times 32$ &
0.542 &
0.289 &
$-1802.835$ 
\\
MOCCa &
$40 \times 40 \times 60$ &
0.653 &
0.290 &
$-1802.969$
\\
\end{tabular}
\end{ruledtabular}
\end{table}

\begin{table}[t]
\caption{\label{tab:pu240_comparison_gs_fi}
Comparison of the calculated superdeformed fission isomer properties of
$^{240}$Pu with the results obtained using
MOCCa~\cite{ryssens_precision}.}
\begin{ruledtabular}
\begin{tabular}{lcccc}
 & 
 $L_x \times L_y \times L_z$ (fm) &
 Step (fm) &
 $\beta_{2}$ &
 $E_{\mathrm{tot}}$ (MeV)
\\
\midrule
This work &
$32 \times 32 \times 32$ &
0.681 &
0.836 &
$-1797.833$
\\
This work &
$40 \times 40 \times 40$ &
0.678 &
0.836 &
$-1797.834$ 
\\
This work &
$32 \times 32 \times 32$ &
0.542 &
0.837 &
$-1797.905$ 
\\
MOCCa &
$40 \times 40 \times 60$ &
0.653 &
0.841 &
$-1798.123$
\\
\end{tabular}
\end{ruledtabular}
\end{table}

\begin{figure}[ht]
  \includegraphics[ width=0.9\columnwidth]{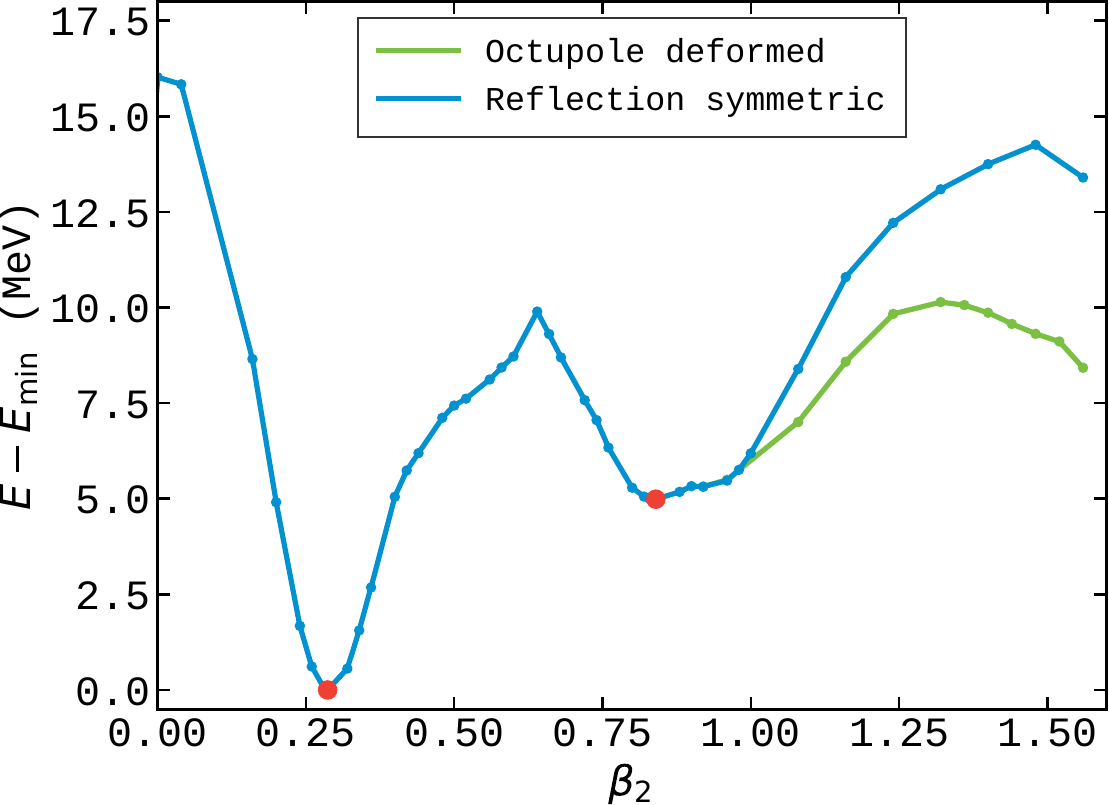}
  \caption{$^{240}$Pu PES as a function of $\beta_2$. Calculations carried out using the SLy4 functional. The unconstrained ground state and fission isomer are marked with red dots.}
  \label{fig:pu240_curve}
\end{figure}

In Fig.~\ref{fig:pu240_curve} the total energy is plotted as a function of the deformation parameter $\beta_2$ for $^{240}$Pu. We performed a reflection symmetric calculation for the whole curve and an octupole calculation for the last part of the curve after the fission isomer. The octupole deformed curve points are computed starting from an initial non-zero $\beta_3$ value. The system is then left free to evolve to the minimal energy configuration in all deformation degrees of freedom except for the constrained $\beta_2$. The octupole configuration is shown to lower the fission barrier energy by several MeV, as well as reducing the $\beta_2$ at which the fission saddle appears. This is a known behaviour and to be expected both in the static and dynamic fission path~\cite{JOHANSSON1961529,pu240_fission}.

\begin{figure}[ht]
  \includegraphics[ width=1.0\columnwidth]{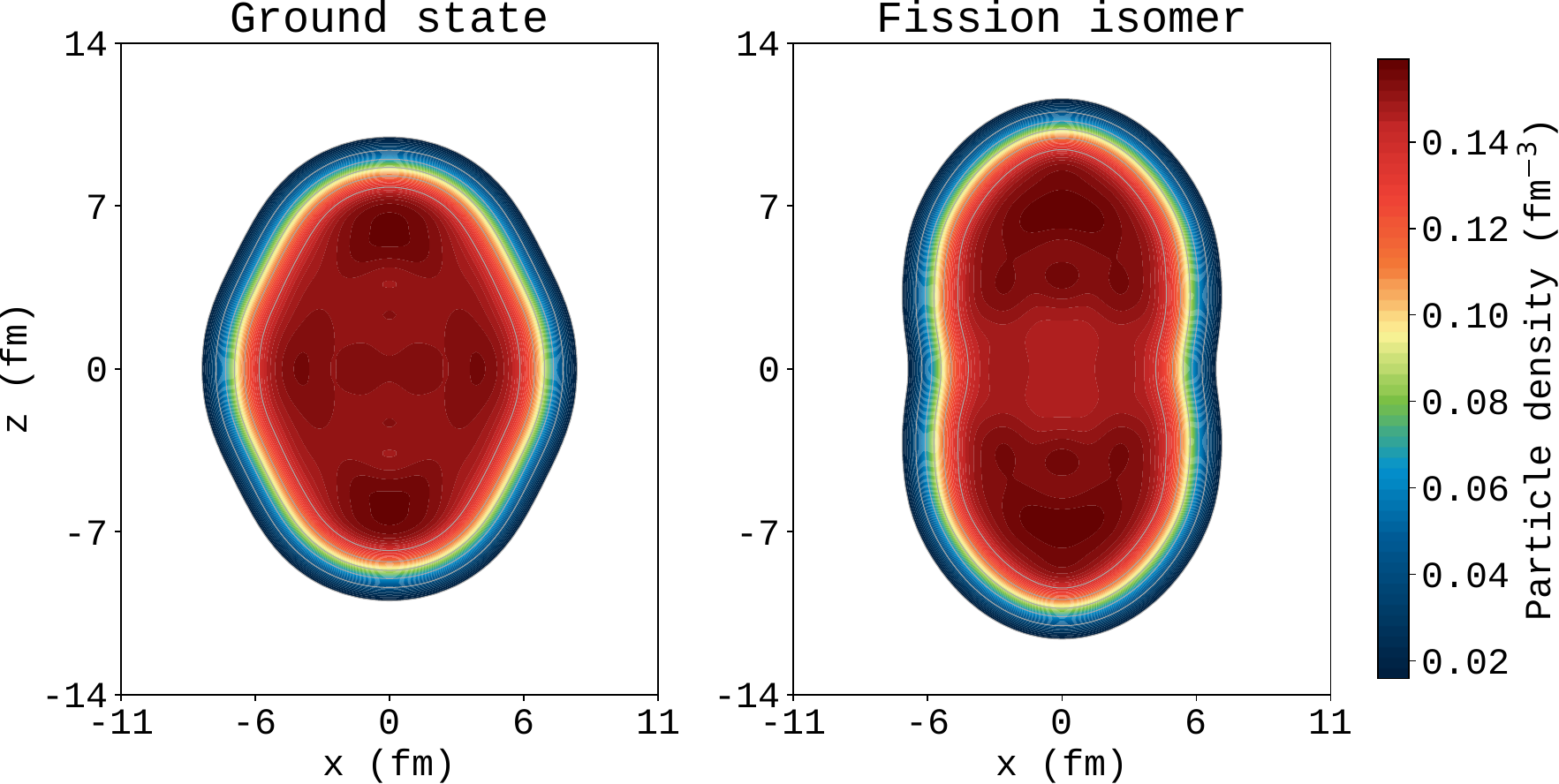}
  \caption{Density contours of the two local minima in $^{240}$Pu, i.e.~the ground state (left) and the superdeformed fission isomer (right). Both densities are reflection symmetric, having $\beta_l=0$ for odd $l$. Some deviation from the pure quadrupole shape is present in both minimum points, the respective $\beta_l$ for even $l$ are shown in Tab.~\ref{tab:pu240_betas}.\label{fig:pu240_densities}}
\end{figure}

In Fig.~\ref{fig:pu240_densities}, the density contours for the two local minima, i.e.~the ground state and the fission isomer are shown. Their respective $\beta_l$ for even $l$ are shown in Tab.~\ref{tab:pu240_betas} to highlight deviations from the pure quadrupole deformation.

\begin{table}[ht]
\caption{Dimensionless deformation parameters of the two local minima in $^{240}$Pu. These values are reported only for the first mesh in Tabs.~\ref{tab:pu240_comparison_gs},\ref{tab:pu240_comparison_gs_fi}.\label{tab:pu240_betas}}
  \centering
\begin{tabular}{lccccc}
\toprule
 & $l=2$ & $l=4$ & $l=6$ & $l=8$ & $l=10$ \\
\midrule
Ground state   & 0.289 & 0.164 & 0.045 & 0.000 & -0.006 \\
Fission isomer & 0.836 & 0.505 & 0.366 & 0.310 & 0.281 \\
\bottomrule
\end{tabular}
\end{table}

\begin{figure}[ht]
  \includegraphics[ width=1.0\columnwidth]{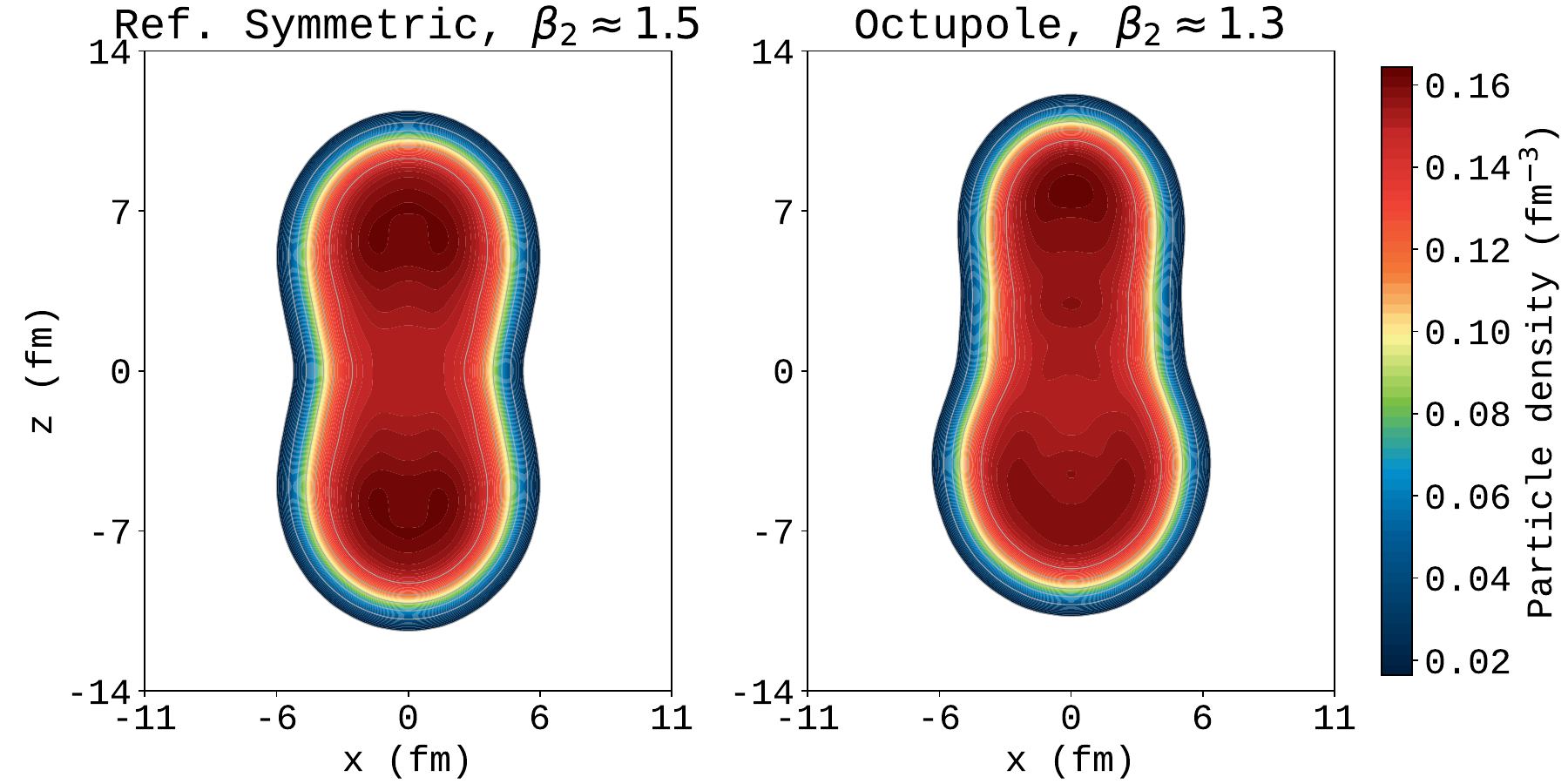}
  \caption{Density contours of the fission saddle in $^{240}$Pu. On the left the reflection symmetric case at $\beta_2\approx 1.5$. On the right, the octupole, reflection asymmetric case at $\beta_2\approx 1.3$. The difference in energy of the two configurations configurations is $\approx 4.2$ MeV.
  \label{fig:pu240_saddle}}
\end{figure}

In Fig.~\ref{fig:pu240_saddle}, the density contours for the first fission saddle in the reflection symmetric and asymmetric cases are shown.

\subsection{\label{sec:ra}$^{224}$Ra octupole deformed ground state}
In this section we study the total energy of $^{224}$Ra, whose ground state is predicted to be in a permanent octupole deformed shape~\cite{Butler2020}.
While the quadrupole deformation has been extensively known and studied, both theory- and experiment-wise, octupole shapes are still under scrutiny. Around the mass region $A\sim 220$ there is an island of nuclei whose ground state octupole deformation has been experimentally identified, among which $^{224}$Ra~\cite{ra_exp_shape}.
Among various consequences of a parity breaking deformation in the ground state, a permanent nuclear dipole moment is induced, which can be used to investigate violations of the CP symmetry~\cite{T_P_odd,Butler2020}. This motivates the interest in studying octupole deformations in the ground state of nuclei.

\begin{figure}[ht]
  \includegraphics[ width=0.9\columnwidth]{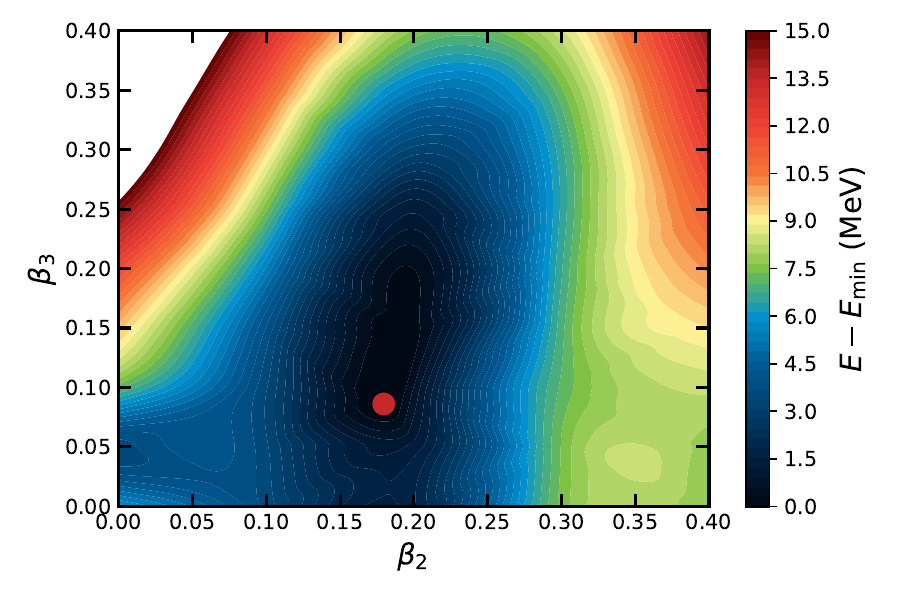}
  \caption{$^{224}$Ra PES as a function of $\beta_2$ and $\beta_3$. Calculations carried out using the SLy4 functional. The unconstrained ground state is marked with a red dot.}
  \label{fig:ra224_map}
\end{figure}

In the following, we study the total energy as a function of $\beta_2$ and $\beta_3$, as shown in Fig.~\ref{fig:ra224_map}. 
The ground state theoretical total energy and deformation parameters are reported in Tab.~\ref{tab:ra224_comparison} and they are compared with the experimental ones, deduced from the $E2$ and $E3$ transitions, assuming the excitation to be rotational~\cite{Butler2020}. A thorough theoretical calculation would require projecting the system on good angular momentum to identify the correct rotational transition. Here, we limit ourselves to showcase the possibility of exploring this deformation on a Cartesian mesh. 

\begin{table}[ht]
\caption{Comparison between the theoretical ground state calculated within the framework of this work and the experimental one.}
\label{tab:ra224_comparison}
  \centering
\begin{tabular}{lccc}
\toprule
Method &
$\beta_{2}$ &
$\beta_{3}$ &
$E_{\mathrm{tot}}$ (MeV)
\\
\midrule

This work &
0.180 &
0.086 &
$-1711.91$
\\

Experiment &
0.223 &
0.122 &
$-1721.96$
\\

\bottomrule
\end{tabular}
\end{table}

\begin{figure}[ht]
  \includegraphics[ width=1.0\columnwidth]{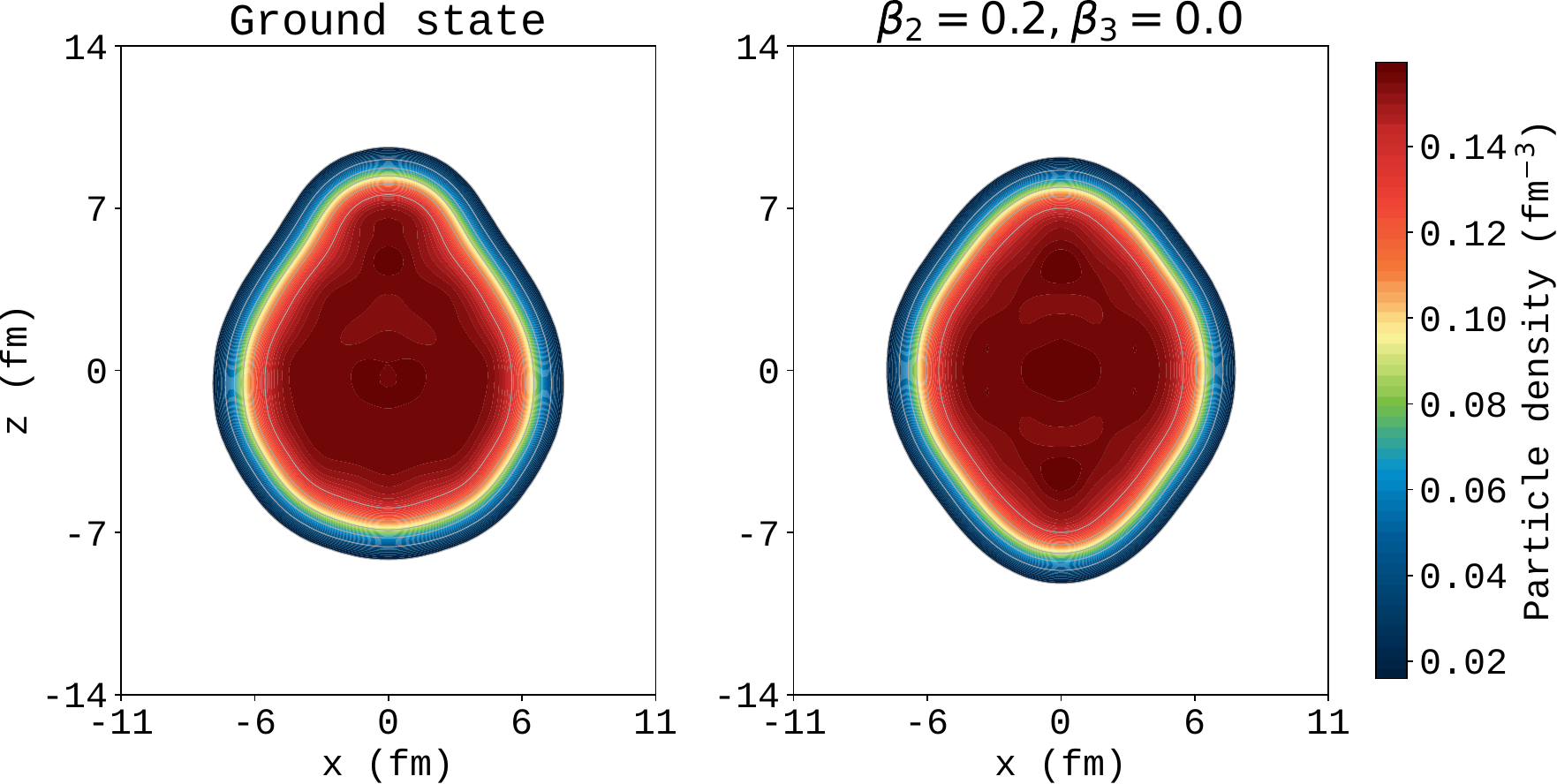}
  \caption{Density contours of the pear-shaped octupole ground state of $^{224}$Ra (left) and a quadrupole configuration (right). Both densities are axially symmetric. The octupole deformation lowers the total energy with respect to the reflection symmetric configuration by $\approx 1.8$ MeV.
  \label{fig:ra224_densities}}
\end{figure}

In Fig.~\ref{fig:ra224_densities}, the density contours for the octupole deformed ground state of $^{224}$Ra and an axial quadrupole configuration are shown. The quadrupole state is chosen with a $\beta_2$ close to the ground state value to highlight the effect of the octupole deformation on the nuclear shape.

\section{Conclusions}
In this article, a new method for the self-consistent solution of the HFB equations based on the GCG method~\cite{gcg1} was proposed.
This method builds upon locally optimal conjugate gradient methods such as LOBPCG~\cite{LOBPCG}, but the preconditioning is done in a different way, that is by using an inverse Hamiltonian step.
This eliminates the need for problem-specific preconditioners and parameter adjusting for optimal convergence.
More importantly, the method was also shown to be more efficient in terms of iterations and CPU time than other algorithms proposed in the literature for 3D lattice space representation.
A code that implements this algorithm has been developed, and its effectiveness was benchmarked against the similar MOCCa code~\cite{ryssens_precision}.

The results obtained in this work represent a further improvement in the efficiency of microscopic self-consistent calculations in Cartesian space.
The computational barrier for systematic applications of Skyrme EDFs in deformed, heavy nuclei is further lowered. Especially for studies that require breaking spatial symmetries, which forbids reducing the matrix size.
Moreover, the method is simple, it does not require manual tuning of any numerical parameter and can be directly applied to similar procedures where large-scale eigenvalue problems need to be solved.
Furthermore, high efficiency will be required for extensions of the method where intensive and/or repetitive solutions of the equations is needed, such as QRPA and angular momentum projection calculations.

\appendix
\section{\label{app:densities}Local densities}
We report here the formulae for the local densities in coordinate space
\begin{align}
    \rho_q(\boldsymbol{r}) =& \sum_{k,\sigma}v_k^2|\psi(\boldsymbol r \sigma)|^2,\\
    \tau_q (\boldsymbol{r})=& \sum_{k,\sigma} v_k^2 |\nabla\psi(\boldsymbol{r}\sigma)|^2,\\
    J_{q, \mu\nu} (\boldsymbol{r})=& \sum_{k,\sigma\sigma'} v_k^2\text{Im}\{{\psi(\boldsymbol{r}\sigma)\partial_\mu\sigma_{\nu,\sigma\sigma'}\psi(\boldsymbol{r}\sigma')}\},
    \\J_{q,\kappa}(\boldsymbol{r})=&\sum_{\mu\nu}\varepsilon_{\kappa\mu\nu}J_{q,\mu\nu}.
\end{align}
Here, the index $k$ runs over the canonical basis wavefunctions. $v_k^2$ denotes the occupation factor of the wavefunction $k$ computed by diagonalizing the density matrix.
\section{\label{app:coulomb}Coulomb field}
The direct part of the Coulomb field, i.e.
\begin{equation}
  U_{c,D} = \frac{e^2}{2}\int d \bm r' \frac{\rho_p(\bm r')}{|\bm r'-\bm r|},
\end{equation}
is computed by solving the Poisson equation
\begin{equation}
  \nabla^2 U_{c, D} = -4\pi e^2\rho_p,
\end{equation}
to avoid the costly evaluation of the integral. The solution is calculated by using a five-point stencil~\cite{finite_diff} for the spatial derivatives, using a quadrupole expansion of the proton density to set Dirichlet boundary conditions
\begin{equation}
  U_{c,D} = \frac{Ze^2}{r} + \frac{\expval{Q_{20}}Y_{20} + 2\expval{Q_{22}}\Re Y_{22}}{r^3}\text{ on }\partial V.
\end{equation}

\section{\label{app:preconditioning}Preconditioning and inverse Hamiltonian}
The aim of preconditioning is to improve the condition number of the problem, i.e.~damp high energy modes that are present at iteration $i$. 
In general, a preconditioner is meant to be a matrix $T$ that approximates the matrix of the problem $h$ while being easier to invert.
In a Schrödinger-like equation a plausible preconditioner to damp high-energy modes could be chosen as
\begin{equation}
  T = \frac{\boldsymbol{p}^2}{2m},
\end{equation}
since high-energy states are dominated by the kinetic energy.
In our case, the main source of off-diagonal terms is the gradient in the momentum operator. This means that the preconditioner we wish to use approximates the problem without improving the sparseness of the the full single-particle matrix $h$, as in Skyrme functionals the only non-local terms are given by laplacians and gradients.
However, note that when using the Fourier transform method, as done in Ref.~\cite{covariant_lobpcg}, the preconditioner is diagonal in momentum space. This may result in a computational speed-up of the preconditioning step.

One possibility to exploit the full information of the matrix is by using the inverse Hamiltonian technique. Assuming to expand the guess orbitals $|\phi\rangle$ in terms of the eigenvalues $\ket{\varphi_\ell}$ of $h$, we can write
\begin{equation}
  \ket{\phi_k} = \sum_\ell c_\ell^k \ket{\varphi_\ell} ,
\end{equation}
where $c_\ell^k = \bra{\varphi_\ell}\ket{\phi_k}$. If we apply the inverse of $h-\varepsilon_0$ to $(\varepsilon_k-\varepsilon_0)\ket{\phi_k}$, we get
\begin{align}
  (h-\varepsilon_0)^{-1}(\varepsilon_k-\varepsilon_0)\ket{\phi_k} =& \sum_\ell c_\ell^k (h-\varepsilon_0)^{-1}(\varepsilon_k-\varepsilon_0)\ket{\varphi_\ell} \nonumber\\=& \sum_\ell c_\ell^k\frac{\varepsilon_k-\varepsilon_0}{\varepsilon_\ell-\varepsilon_0} \ket{\varphi_\ell}.
  \label{eq:inv}
\end{align}
From the last step of Eq.~\eqref{eq:inv} it is clear that the contribution of high energy eigenvectors is damped at each inverse step, while the correct contribution $\varepsilon_k \approx \varepsilon_\ell$ remains unaffected.
Shifting $h$ by $\varepsilon_0$ has two effects: (a) it shifts the spectrum of interest to positive eigenvalues, otherwise the method would not be well-defined for negative energy states, as the quantity $|\varepsilon_k/\varepsilon_\ell|$ diverges for $\varepsilon_\ell \sim 0$.~(b) the shift allows to numerically avoid singularities by choosing a small quantity $c > 0$ to be added such that all eigenvalues of $h-\varepsilon_0 + c$ are strictly positive.
Solving for the inverse of $h$ on the guess orbitals has a clear computational time advantage. After some iterations, the orbitals will be very close to the real eigenvectors of the matrix, meaning that an iterative solution of $h^{-1}\ket\phi$ will be found by the chosen algorithms in less CPU time as the iterations converge.


\nocite{*}

\bibliography{bibliography}

\end{document}